\documentclass[runningheads]{llncs}
\usepackage[T1]{fontenc}
\usepackage{graphicx,verbatim}
\usepackage{amssymb}
\usepackage{latexsym}
\usepackage{amsmath}
\usepackage{graphicx}
\usepackage{textcomp}
\usepackage{booktabs}
\usepackage{color}
\usepackage{multirow}
\usepackage{array}
\usepackage{multicol}
\usepackage{layouts}
\usepackage{braket}
\usepackage{siunitx}
\usepackage{subfigure}
\usepackage{float}
\usepackage{hyperref}
\usepackage{bm}
\begin{document}
\title{Bridging Classical and Learning-based Iterative Registration through Deep Equilibrium Models}
%
\begin{comment}  %% Removed for anonymized MICCAI 2025 submission
\author{First Author\inst{1}\orcidID{0000-1111-2222-3333} \and
Second Author\inst{2,3}\orcidID{1111-2222-3333-4444} \and
Third Author\inst{3}\orcidID{2222--3333-4444-5555}}
%
\authorrunning{F. Author et al.}
\institute{Princeton University, Princeton NJ 08544, USA \and
Springer Heidelberg, Tiergartenstr. 17, 69121 Heidelberg, Germany
\email{lncs@springer.com}\\
\url{http://www.springer.com/gp/computer-science/lncs} \and
ABC Institute, Rupert-Karls-University Heidelberg, Heidelberg, Germany\\
\email{\{abc,lncs\}@uni-heidelberg.de}}

\end{comment}

 %% Removed for anonymized MICCAI 2025 submission
\author{Yi Zhang \and
Yidong Zhao \and
Qian Tao}
\authorrunning{Y. Zhang et al.}
\institute{Department of Imaging Physics, Delft University of Technology, the Netherlands \email{\{y.zhang-43,y.zhao-8,q.tao\}@tudelft.nl}}
\maketitle              % typeset the header of the contribution
\begin{abstract}
Deformable medical image registration is traditionally formulated as an optimization problem. While classical methods solve this problem iteratively, recent learning-based approaches use recurrent neural networks (RNNs) to mimic this process by unrolling the prediction of deformation fields in a fixed number of steps. However, classical methods typically converge after sufficient iterations, but learning-based unrolling methods lack a theoretical convergence guarantee and show instability empirically. In addition, unrolling methods have a practical bottleneck at training time: GPU memory usage grows linearly with the unrolling steps due to backpropagation through time (BPTT). To address both theoretical and practical challenges, we propose DEQReg, a novel registration framework based on Deep Equilibrium Models (DEQ), which formulates registration as an equilibrium-seeking problem, establishing a natural connection between classical optimization and learning-based unrolling methods. DEQReg maintains constant memory usage, enabling theoretically unlimited iteration steps. Through extensive evaluation on the public brain MRI and lung CT datasets, we show that DEQReg can achieve competitive registration performance, while substantially reducing memory consumption compared to state-of-the-art unrolling methods. We also reveal an intriguing phenomenon: the performance of existing unrolling methods first increases slightly then degrades irreversibly when the inference steps go beyond the training configuration. In contrast, DEQReg achieves stable convergence with its inbuilt equilibrium-seeking mechanism, bridging the gap between classical optimization-based and modern learning-based registration methods.
% Authors must provide keywords and are not allowed to remove this Keyword section.

\end{abstract}
\section{Introduction}
Deformable image registration (DIR) is a fundamental task in medical image analysis that aims to discover the spatial correspondence between pairs of images \cite{staring2009registration,sotiras2013deformable,haskins2020deep}. DIR can be formulated as an optimization problem to find a parameterized deformation field that aligns the images. The objective typically comprises two terms: image similarity for alignment accuracy, and regularization for deformation smoothness and anatomical plausibility.  Historically, this optimization problem has been solved by iterative algorithms such as adaptive gradient descent \cite{ashburner2007fast,klein2009elastix,avants2011reproducible}. While successful, the iterative nature of these methods leads to long computation time seeking convergence, limiting their real-time clinical applications. With recent advances in deep learning (DL), data-driven approaches have emerged as a promising alternative, by directly inferring the transformation field in a single forward-pass through parametrized neural networks \cite{de2019deep,balakrishnan2019voxelmorph}. More recently, inspired by classical iterative methods, unrolling methods \cite{hering2019mlvirnet,zhao2019recursive,mok2020large,falta2022learning,qiu2022embedding,zhang2024recurrent} were proposed to learn the optimization process in multiple steps. Compared to single-step inference, these unrolling methods demonstrated further improved performance \cite{zhao2019recursive,falta2022learning,qiu2022embedding}.
\begin{figure}[!t]
    \centering
    \includegraphics[width=1\linewidth]{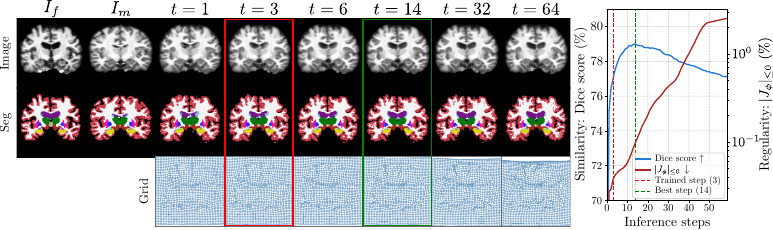}
    \caption{Performance of a learning-based unrolling method (3 resolution level, 3 steps per level, 9 steps in total). Inference steps $t$ denotes steps on the highest level.}
    \label{fig1} 
\end{figure}  

Despite improved accuracy and generalizability over single-step inference, a fundamental gap exists between learning-based unrolling methods and classical optimization methods. While classical optimization methods pursue convergence for optimal performance and often require >100 iterations, current unrolling methods rely on a limited number of steps (<10) at both training and inference time. At training time, this limitation stems from the reliance on backpropagation through time (BPTT) \cite{rnn}, which causes a linearly growing memory footprint that quickly surpasses the practical GPU capacity. At inference time, extended iterations can be made beyond the training steps, but existing implementations maintain the same iteration steps in both phases. This raises a curious question as to what would happen when inference steps go beyond training ones, which we show in Fig. \ref{fig1}. It can be observed that extended inference steps initially improve the Dice score and then degrade it, while the deformation regularity expressed by percentage of negative Jacobian determinants of deformation field (folded voxels) deteriorates consistently. The empirical results show that learning-based unrolling methods lack a convergence guarantee, which is well-established in classical methods and highly desirable for DIR.
 
To bridge this conceptual and empirical gap, we propose DEQReg, which seeks a converged estimation of registration fields based on recent Deep Equilibrium Models (DEQ) \cite{bai2019deep}. DEQReg is the first attempt to formulate the entire learning-based registration process as an end-to-end unsupervised fixed-point problem. It offers both theoretical and practical advantages: first, by explicitly seeking equilibrium points during iterative registration, it establishes a natural connection between classical optimization and learning-based unrolling methods. Second, DEQ's implicit differentiation \cite{bai2019deep,krantz2002implicit} enables potentially unlimited iteration steps with constant memory usage, bypassing the memory limitations of BPTT. Unlike recent works that apply DEQ as a post-processing step for pre-trained models \cite{hu2023plug} or as a solver for supervised regression and inverse problems \cite{bai2022deep,gilton2021deep}, our work presents two new contributions for DIR:
\begin{itemize}
    \item We propose DEQReg, the first end-to-end unsupervised registration framework that formulates DIR as a fixed-point equilibrium problem. Extensive experiments on the public brain MRI and lung CT datasets demonstrate that our approach achieves competitive accuracy with constant memory complexity through implicit differentiation.
    \item We empirically demonstrate that DEQReg achieves stable convergence at inference time while revealing an intriguing phenomenon in current unrolling methods: Their performances are bound to deteriorate after initial slight improvement beyond the predefined training step, highlighting the significance of our principled equilibrium-seeking design.
\end{itemize}
\section{Methods}
\subsection{Iterative Learning-based Image Registration via Unrolling}
Given a fixed image $I_f$ and a moving image $I_m$, DIR aims to find a dense displacement field $\mathbf{u}$ that aligns $I_m$ with $I_f$. The deformation field ${\phi}$ is defined as ${\phi}(\mathbf{x}) = \mathbf{x} + \mathbf{u}(\mathbf{x})$ for any spatial coordinate $\mathbf{x}$. For notational simplicity, we use ${\phi}$ hereafter. DIR can be formulated as an optimization problem:
\begin{equation}
    {\phi}^* = \underset{{\phi}}{\operatorname{argmin}} \, \mathcal{L}_\text{sim}(I_f, I_m \circ {\phi}) + \lambda \mathcal{L}_\text{reg}({\phi}),
\end{equation}
where $\mathcal{L}_{\text{sim}}$ measures image similarity, $\mathcal{L}_{\text{reg}}$ enforces smoothness of the deformation, and $\lambda$ balances these competing terms. Classical methods typically solve this optimization through gradient descent:
\begin{equation}
    {\phi}_{t+1} = \phi_t - \eta_t \nabla_{\phi} (\mathcal{L}_\text{sim}(I_f, I_m \circ {\phi}_t) + \lambda \mathcal{L}_\text{reg}({\phi}_t)), \quad t = 0,1,\ldots,L,
\end{equation}
where $\eta_t$ denotes an adaptive step size. The optimization typically continues until convergence or reaches a maximum iteration count $L$, which often exceeds 100 steps for sufficient optimization.
\begin{figure}[!ht]
    \centering
    \includegraphics[width=1\linewidth]{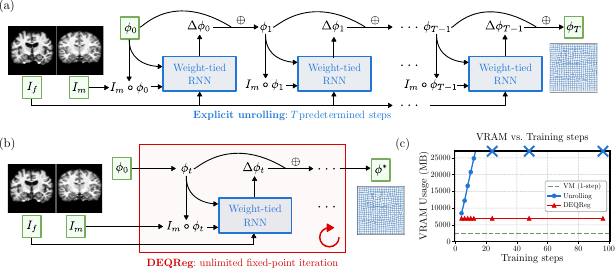}
    \caption{(a) Illustration of a typical learning-based registration method that uses explicit unrolling of $T$ steps in training. (b) DEQReg uses implicit fixed-point iteration, enabling theoretically unlimited steps in training. (c) Memory usage when training on a 3D brain MRI pair, where blue crosses indicate memory overflow for a 24GB GPU.}
    \label{fig:method}
\end{figure}
In iterative learning-based registration, this optimization is often reformulated by unrolling into a fixed number of steps $T$. The update rule is modeled using a weight-tied RNN as shown in Fig. \ref{fig:method}, where parameters $\theta$ are shared across steps to learn the optimal update function $f_\theta$:
\begin{equation}
    {\phi}_{t+1} = \phi_t + f_\theta(I_f, I_m, {\phi}_t), \quad t = 0,1,\ldots,T-1.
\end{equation}
The update function $f_\theta$ is learned by minimizing the expected loss over a training dataset $\mathcal{D} = \{(I_f^i, I_m^i)\}_{i=1}^M$:
\begin{equation}
    \min_{\theta} \mathbb{E}_{(I_f, I_m) \sim \mathcal{D}} [\mathcal{L}(I_f, I_m, \phi_T) + \sum_{t=1}^{T} w_t \mathcal{L}(I_f, I_m, \phi_t)]. \label{eq:obj}
\end{equation}
Training such unrolling networks through BPTT \cite{rnn} requires storing \emph{all} intermediate states, leading to $O(T)$ memory usage and typically restricting steps to $T < 10$ due to hardware constraints.

\subsection{Deep Equilibrium Models for Deformable Image Registration}
To overcome the limitations of unrolling approaches that can only allow for a fixed number of iterations, we reformulate the iterative process using the Deep Equilibrium Model (DEQ) \cite{bai2019deep} to find the equilibrium of a fixed-point equation. In its general form, a fixed point $\mathbf{x}^*$ of a function $g$ satisfies $\mathbf{x}^* = g(\mathbf{x}^*)$. Here, we design our RNN $f_\theta$ to learn residual updates, where DEQ seeks an equilibrium solution ${\phi}^*$ through residual updates that satisfy this fixed-point condition:
\begin{equation}
    {\phi}^* = g_\theta(I_f,I_m,{\phi}^*) = {\phi}^* + f_\theta(I_f, I_m, {\phi}^*),
\end{equation}
which is equivalent to finding ${\phi}^*$ where $f_\theta(I_f, I_m, {\phi}^*) = 0$. This formulation provides a principled way to enforce convergence through vanishing updates, analogous to classical optimization but with the learned update by $f_\theta$.

Unlike BPTT that requires storing and backpropagating throughout the forward trajectory, a key advantage of DEQ is its ability to compute derivatives directly from the derived equilibrium state. When the system converges to a fixed point ${\phi}^*$ where $f_\theta(I_f, I_m, {\phi}^*) = 0$, we can leverage the implicit function theorem (IFT) \cite{bai2019deep,krantz2002implicit} to compute gradients without tracking intermediate states:

\begin{equation}
    \frac{\partial \mathcal{L}}{\partial \theta} =\frac{\partial \mathcal{L}}{\partial {\phi}^*} \left(I - \frac{\partial g_\theta}{\partial {\phi}^*}\right)^{-1} \frac{\partial g_\theta(I_f,I_m,{\phi}^*)}{\partial \theta}, \label{eq:ift}
\end{equation}
where $I$ is an identity matrix of the same size as $\frac{\partial g_\theta}{\partial {\phi}^*}$. Leveraging IFT, equilibrium points can be derived using any black-box root-finding algorithm, e.g. Broyden's method \cite{broyden1965class}. Fig. \ref{fig:method} provides a schematic comparison between DEQ and existing unrolling methods. By focusing on the equilibrium state rather than the iteration trajectory, DEQ achieves constant memory complexity while enabling theoretically unlimited iterations, with convergence design (i.e. fixed point condition).

\subsection{Training the DEQ}
Despite the memory efficiency of DEQ, directly computing the inverse Jacobian term $(I - \frac{\partial g_\theta}{\partial {\phi}^*})^{-1}$ in Eq. \eqref{eq:ift} remains computationally intensive. To address this challenge, we use the phantom gradient approach \cite{geng2021training,geng2023torchdeq}, which offers a structured approximation of IFT that preserves the same descent direction while being computationally efficient. The key idea is to approximate the inverse Jacobian through a sequence of damped iterations. Starting from the converged fixed point ${\phi}^0 = {\phi}^*$, we generate this sequence with a damping factor $\tau \in (0,1)$:

\begin{equation}
    {\phi}^{p+1} = \tau g_\theta(I_f, I_m, {\phi}^p) + (1-\tau){\phi}^p, \quad p = 0,\ldots,K-2.
\end{equation}
Using the generated sequence $\{{\phi}^p\}_{p=0}^{K-1}$, we can approximate the gradient as:

\begin{equation}
    \frac{\partial \mathcal{L}}{\partial \theta} \approx \frac{\partial \mathcal{L}}{\partial {\phi}^*}\mathbf{A}, \quad \text{where} \quad \mathbf{A} = \tau \sum_{k=0}^{K-1} \prod_{s=k+1}^{K-1} \left(\tau \frac{\partial g_\theta}{\partial {\phi}}\bigg|_{{\phi}^s} + (1-\tau)I\right) \frac{\partial g_\theta}{\partial \theta}\bigg|_{{\phi}^k}. \label{eq:pg}
\end{equation}
This truncated gradient converges to the exact implicit gradient in Eq. \ref{eq:ift} as $K \to \infty$ under mild conditions (see \cite{geng2021training} for detailed theoretical analysis). 

Additionally, we can leverage intermediate state correction during the forward process to regularize the iteration behavior \cite{falta2022learning,bai2022deep,zhang2024recurrent} in training. Given a training trajectory $[{\phi}_0, {\phi}_1, \ldots, {\phi}_N, {\phi}^*]$, we sample $S$ intermediate states with equal intervals, where $\mathcal{S} \subset \{0,1,\ldots,N\}$ with $|\mathcal{S}| = S$, to compose the total loss:
\begin{equation}
    \mathcal{L}_{\text{total}} = \mathcal{L}(I_f, I_m, {\phi}^*) + \gamma\sum_{t \in \mathcal{S}} \mathcal{L}(I_f, I_m, {\phi}_t), \quad \gamma>0, \label{eq:deq_train}
\end{equation}
The intermediate state gradients are computed via phantom gradients, enabling effective training of the equilibrium-seeking network $f_\theta$.
\section{Experiments}
\noindent \textbf{Datasets} We evaluated our method on two public datasets: (1) \textbf{OASIS} \cite{marcus2007open}, a brain MRI dataset containing 414 T1-weighted scans for inter-subject registration. Following \cite{hoopes2021hypermorph}, scans were preprocessed and resampled to a size of $128 \times 128 \times 128$, with 30 anatomical segmentation labels (300/30/84 for train/val/test). (2) \textbf{NLST} \cite{aberle2011reduced}, a lung CT dataset from Learn2Reg challenge \cite{hering2022learn2reg} with 210 intra-subject inhale-exhale pairs. The affinely prealigned scans are of size $224 \times 192 \times 224$, with keypoint annotations and lung masks (170/10/30 for train/val/test). 

\noindent \textbf{Evaluation Metrics} We evaluate both anatomical alignment and deformation smoothness. For alignment, we compute average Dice score and Hausdorff Distance (HD) on OASIS using the structures in \cite{hoopes2021hypermorph}, and keypoint target registration error (TRE) and lung Dice score on NLST. For deformation smoothness, we compute two Jacobian-based metrics ($J_{{\phi}} = \nabla {\phi}$): the percentage of non-diffeomorphic (folded) voxels $|J_{\phi}|_{\leq 0}$ and the standard deviation of log-Jacobian determinant $\text{std}(\log \vert J_{{\phi}} \vert)$ quantifying the spatial consistency of volume changes.

\noindent \textbf{Baseline Methods} We compare our method against the following approaches: (1) \textbf{elastix} \cite{klein2009elastix}, a widely-used classical registration method, configured with 3-level B-spline transformation and grid spacing of 4 voxels; (2) \textbf{VoxelMorph} \cite{balakrishnan2019voxelmorph}, a well-established one-step DL method; and two state-of-the-art unrolling methods: (3) \textbf{GraDIRN} \cite{qiu2022embedding}, which integrates explicit similarity gradients alongside neural network predictions in 3 resolution levels with $T=3$ per level ($w_t=0$), and (4) \textbf{RIIR} \cite{zhang2024recurrent} with single resolution and $T=6$ iterations using exponentially increasing weights ($w_t = 10^{\frac{t-1}{T-1}}$). Our proposed DEQReg employs a fixed-point iteration strategy with maximum $T=48$ steps with $S=3$ and $\gamma=0.5$ in Eq. \ref{eq:deq_train}.

\noindent \textbf{Convergence Analysis Setup} In this study, we investigate how DL-based unrolling methods behave when varying inference steps while keeping training steps fixed. Besides GraDIRN and RIIR, we evaluate two additional variants: (1) Unrolling (the vanilla version of GraDIRN without incorporating similarity gradients) and (2) DEQReg-12, DEQReg with $T=12$ and $S=0$. For GraDIRN and Unrolling, we vary the steps at the highest resolution.

\noindent \textbf{Implementation Details} For fair comparison, we use the same U-Net \cite{ronneberger2015unet} configuration as VoxelMorph and RIIR for DEQReg. For each step $t$, our network takes concatenated $[I_f, I_m \circ \phi_t, \phi_t]$ as input into the neural network, consistent with other iterative methods. The loss combines local normalized cross correlation (LNCC) with window size 5 as $\mathcal{L}_{\text{sim}}$ and diffusion regularization as $\mathcal{L}_{\text{reg}}$, with $\lambda = 0.1$ (OASIS) and 0.125 (NLST). Models were trained using AdamW (lr=$1\times 10^{-4}$) for 100 epochs. Code and pretrained weights will be made publicly available.

\section{Results and Discussions}
\begin{table}[!htbp]
    \caption{Quantitative comparison of registration methods. Mem. denotes GPU memory usage and Time shows training/inference time per pair. Best performance is in \textbf{bold} and second best is \underline{underlined}, with $^*$ denotes statistical significance ($p < 0.05$, Wilcoxon).}
    \label{tab:summary}
    \centering
    \fontsize{8pt}{9.6pt}\selectfont
    \begin{tabular}{l cccccc}
    \toprule  
    \textbf{OASIS} & Dice (\%)↑ & HD (mm)↓ & $|J_{\phi}|_{\leq 0}$ (\%)↓ & std$(\log|J_{\phi}|)$↓ & Mem. (MB) & Time (s) \\
    \midrule
    Affine & 61.2(7.3) & 8.56(1.34) & - & - & - & - \\
    elastix & 75.2(3.8) & 2.89(0.58) & 0.068(0.041) & \underline{0.551}(0.134) & - & 35.45 \\
    VoxelMorph & 77.8(3.7) & 2.74(0.49) & 0.063(0.031) & 0.601(0.121) & 2645 & 0.20/0.14 \\
    GraDIRN & 80.3(3.2) & 2.47(0.41) & 0.071(0.029) & 0.568(0.114) & 16619 & 0.59/0.24 \\
    RIIR & \underline{80.6}(3.1) & \textbf{2.43}(0.39) & \underline{0.058}(0.031) & 0.556(0.118) & 12542 & 0.60/0.30 \\
    DEQReg & \textbf{80.7}$^*$(3.1) & \underline{2.43}(0.40)& \textbf{0.053}$^*$(0.033) & \textbf{0.542}$^*$(0.123) & 6925 & 1.09/0.98 \\
    \midrule
    \textbf{NLST} & TRE (mm)↓ & Dice (\%)↑ & $|J_{\phi}|_{\leq 0}$ (\%)↓ & std$(\log|J_{\phi}|)$↓ & Mem. (MB) & Time (s) \\
    \midrule
    Affine & 8.43(3.97) & 87.3(4.1) & - & - & - & - \\
    elastix & 3.24(1.74) & 94.8(0.7) & \underline{0.110}(0.136) & 0.636(0.370) & - & 32.51 \\
    VoxelMorph & 4.08(2.31) & 95.6(1.2) & 0.143(0.290) & 0.671(0.506) & 1522 & 0.24/0.12 \\
    GraDIRN & 2.26(1.87) & 96.5(1.2) & 0.165(0.278) & 0.738(0.512) & 9313 & 0.41/0.18 \\
    RIIR & \textbf{2.21}$^*$(1.78) & \underline{96.5}(1.1) & 0.120(0.250) & \underline{0.598}(0.481) & 7029 & 0.40/0.16 \\
    DEQReg & \underline{2.23}(2.11) & \textbf{96.6}(1.5) & \textbf{0.079}$^*$(0.078) & \textbf{0.586}$^*$(0.290) & 3983 & 0.74/0.53 \\
    \bottomrule 
    \end{tabular}
\end{table}
\noindent \textbf{Main Results} Table \ref{tab:summary} presents a comprehensive comparison of our DEQReg against classical and learning-based approaches, with visual results shown in Fig. \ref{fig:qualitative}. DEQReg achieves a statistically significant improvement in Dice scores while maintaining competitive HD on OASIS, and demonstrates comparable performance in TRE and Dice score on NLST. Notably, by decomposing deformation into more steps, DEQReg achieves the lowest percentage of $|J_{{\phi}}|_{\leq 0}$ and $\text{std}(\log|J_{{\phi}}|)$, indicating superior deformation regularity. Remarkably, DEQReg substantially reduces memory consumption compared to other DL methods. %These results confirm that DEQReg's equilibrium-based design leads to memory-efficient registration with improved deformation regularity given the same network backbone.

\begin{figure}[!b]
    \centering
    \includegraphics[width=1\linewidth]{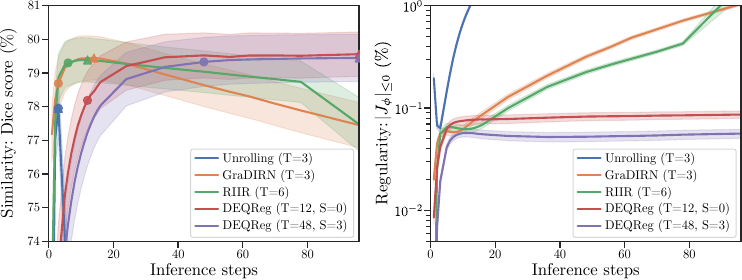}
    \caption{Convergence analysis for different registration methods. Left: Dice score (\%) where \(\bullet\) indicates the results at the trained step count and \(\blacktriangle\) marks the steps achieving best Dice performance; Right: negative Jacobian determinant ratio (\%). Shaded areas show standard deviation. DEQReg exhibits convergence beyond training steps.}
    \label{fig:convergence}
\end{figure}

\noindent \textbf{Convergence Analysis} Fig. \ref{fig:convergence} and Table \ref{tab:validation_results} present the convergence behavior of different methods. The vanilla unrolling method shows a severe performance drop beyond its trained steps. Although GraDIRN and RIIR show modest improvement when inference steps exceed training configuration due to their image similarity-driven updates, they eventually suffer from deteriorating deformation regularity (shown in Fig. \ref{fig:qualitative}, row 5) and Dice. In contrast, both DEQReg variants demonstrate stable convergence. 
Notably, DEQReg shows potential for improvement with increased iterations, as its optimal Dice occur at maximum steps. Comparing the gap between trained-step and best-achievable performance, methods with intermediate state correction as in Eq. \ref{eq:deq_train} (RIIR and DEQReg) achieve near-optimal results at trained steps, confirming their effectiveness. In particular, the full DEQReg consistently maintains superior deformation regularity while achieving competitive accuracy, as shown in Fig. \ref{fig:convergence}.
\begin{table}[!t] 
\centering
\caption{Comparison of registration performance at trained steps versus best achievable steps on validation set. Best performance is in \textbf{bold} and second best is \underline{underlined}.}
\fontsize{8pt}{9.6pt}\selectfont
%\small
\begin{tabular}{@{}l cc ccc c@{}}
\toprule

& \multicolumn{3}{c}{Trained Steps} & \multicolumn{3}{c}{Best Steps} \\
\cmidrule(lr){2-4} \cmidrule(lr){5-7}
Method & Steps & Dice (\%) & $|J_{\phi}|_{\leq 0}$ (\%) & Steps & Dice (\%) & $|J_{\phi}|_{\leq 0}$ (\%) \\
\midrule
Unrolling & 3 & 77.95(4.67) & 0.064(0.031) & 3 & 77.95(4.67) & 0.064(0.031) \\
GraDIRN & 3 & 78.69(4.20) & \underline{0.057}(0.029) & 14 & 79.44(3.57) & 0.078(0.033) \\
RIIR & 6 & 79.30(3.72) & 0.066(0.031) & 12 & 79.39(3.58) & \underline{0.062}(0.028) \\
DEQReg-12 & 12 & 78.19(4.68) & 0.077(0.041) & 96 & \textbf{79.53}(3.62) & 0.087(0.039) \\
DEQReg & 48 & \textbf{79.33}(3.97) & \textbf{0.053}(0.030) & 96 & 79.45(3.84) & \textbf{0.056}(0.031) \\
\bottomrule
\end{tabular}
\label{tab:validation_results}
\end{table}

\begin{figure}[!tbp]
    \centering
    \includegraphics[width=1\linewidth]{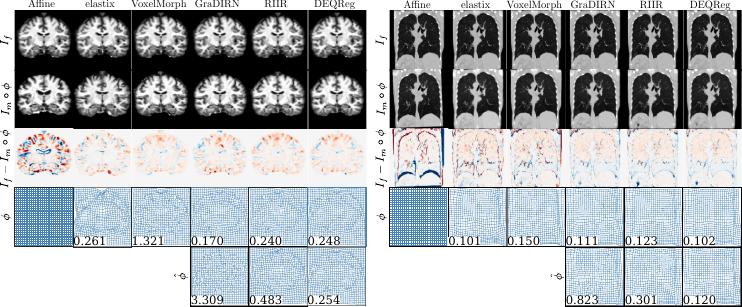}
    \caption{Qualitative comparison of registration methods on brain MRI and lung CT images. In the last row, all unrolling methods go to the excessive step count of $T=96$ to evaluate the convergence of $\phi$ to $\hat{\phi}$. The percentage of $|J_{\phi}|_{\leq 0}$ is reported at the bottom-left corner to indicate deformation regularity.}
    \label{fig:qualitative} 
\end{figure}
\section{Conclusion}
For deformable medical image registration, classical optimization methods typically converge, but modern learning-based unrolling methods lack the theoretical guarantee of convergence and suffer from memory bottlenecks. This creates a gap between the principled convergence of classical methods and the lack of convergence in learning-based methods. In this work, we propose DEQReg to bridge this gap by reformulating learning-based registration as an equilibrium-seeking problem. Our experiments show that this reformulation enables stable convergence, in contrast to existing unrolling methods that tend to degrade when further unrolled beyond the training steps. DEQReg achieves competitive accuracy and improved regularity with constant memory usage, effectively unifying classical optimization principles and learning-based advantages in speed and flexibility.

\begin{credits}
\subsubsection{\ackname} We gratefully acknowledge the Dutch Research Council (NWO) and Amazon Science for financial and computing support.

\subsubsection{\discintname}
The authors have no competing interests to declare that are relevant to the content of this article.
\end{credits}

\begin{comment}  %% removed for anonymized MICCAI 2025 submission. 
    
    % The following acknowledgement and disclaimer sections should be removed for the double-blind review process.  
    % If and when your paper is accepted, reinsert the acknowledgement and the disclaimer clause in your final camera-ready version.

\begin{credits}
\subsubsection{\ackname} A bold run-in heading in small font size at the end of the paper is
used for general acknowledgments, for example: This study was funded
by X (grant number Y).

\subsubsection{\discintname}
It is now necessary to declare any competing interests or to specifically
state that the authors have no competing interests. Please place the
statement with a bold run-in heading in small font size beneath the
(optional) acknowledgments\footnote{If EquinOCS, our proceedings submission
system, is used, then the disclaimer can be provided directly in the system.},
for example: The authors have no competing interests to declare that are
relevant to the content of this article. Or: Author A has received research
grants from Company W. Author B has received a speaker honorarium from
Company X and owns stock in Company Y. Author C is a member of committee Z.
\end{credits}

\end{comment}

\bibliographystyle{splncs04}
\bibliography{sample}

\end{document}